\documentclass[%
 reprint,
superscriptaddress,
longbibliography,
 amsmath,amssymb,
 aps,
 twocolumn,
 floatfix,
 aps, 
 showkeys, prl
]{revtex4-2}

\usepackage{graphicx}% Include figure files
\usepackage{dcolumn}% Align table columns on decimal point
\usepackage{bm}% bold math
\usepackage[unicode,colorlinks=true,allcolors=blue]{hyperref} % colored references and citations
\usepackage{siunitx}
\usepackage{here}
\usepackage{amssymb}
\usepackage[dvipsnames]{xcolor}
\usepackage{color}
\usepackage{pifont} % Required for dingbat symbols
\usepackage{wasysym}
\usepackage{stackengine}
\usepackage{pdfrender}
\newcommand{\thickredstar}{%
  \textpdfrender{
    TextRenderingMode=Stroke,
    StrokeColor=red,
    LineWidth=0.5pt, % Lower this for a thinner line, raise it for thicker
  }{\ding{73}}%
}
\newcommand{\thickbluestar}{%
  \textpdfrender{
    TextRenderingMode=Stroke,
    StrokeColor=RoyalBlue,
    LineWidth=0.5pt, % Lower this for a thinner line, raise it for thicker
  }{\ding{73}}%
}

\definecolor{darkgreen}{RGB}{0,80,0}

\DeclareSIUnit\ecm{\text{\textit{e}}cm}

\definecolor{mplblue}{HTML}{1f77b4}
\definecolor{mplorange}{HTML}{ff7f0e}
\definecolor{mplgreen}{HTML}{2ca02c}

\begin{document}
%TC:ignore
\preprint{APS/123-QED}

\title{%A New High-Intensity Frontier for Ultracold Neutrons\\ or \\ 
A New High-Intensity Source for Ultracold Neutrons}% Force line breaks with \\
%\thanks{A footnote to the article title}%
\author{K.~Abe}
\affiliation{Nagoya University, Nagoya, Aichi, Japan}
\author{S.~Ahmed}
\affiliation{University of Manitoba, Winnipeg, MB, Canada}
\author{B.~Algohi}
\affiliation{University of Manitoba, Winnipeg, MB, Canada}
\author{D.~Anthony} %\altaffiliation{Present address: Queen's University, Kingston, Ontario, Canada} % ugrad 2024 
\affiliation{TRIUMF, Vancouver, BC, Canada}
\author{L.~Barr\'on-Palos}
\affiliation{Instituto de F\'isica, Universidad Nacional Aut\'onoma de M\'exico, Mexico City, Mexico}
%\affiliation{Instituto de F\'sica, Universidad Nacional Aut\'onoma de M\'exico, Apartado Postal 20-364, 01000, Mexico}
\author{Y.~Bylinsky}
\affiliation{TRIUMF, Vancouver, BC, Canada}
\author{M.~Boss\'e} % ugrad 2025
\affiliation{TRIUMF, Vancouver, BC, Canada}
\author{M.P.~Bradley}
\affiliation{University of Saskatchewan, Saskatoon, SK, Canada}
\author{A.~Brossard}
\affiliation{TRIUMF, Vancouver, BC, Canada}
%\author{T.~Bui} % ugrad 2024 Winnipeg, hasn't worked on source
%\affiliation{University of Manitoba, Winnipeg, MB, Canada}
\author{J.~Chak}
\affiliation{TRIUMF, Vancouver, BC, Canada}
\author{R.~Chiba}
\affiliation{Simon Fraser University, Burnaby, BC, Canada}
\author{C.~Davis}
\affiliation{TRIUMF, Vancouver, BC, Canada}
\author{R.~de Vries} % ugrad 2024-2025
\affiliation{The University of Winnipeg, Winnipeg, MB, Canada}
\author{K.~Dong}    % ugrad 2025-2026, plot maker for this paper
\affiliation{TRIUMF, Vancouver, BC, Canada}
\author{K.~Drury} % ugrad 2024-2025
\affiliation{TRIUMF, Vancouver, BC, Canada}
\author{B.~Franke}
\affiliation{TRIUMF, Vancouver, BC, Canada}
\affiliation{The University of British Columbia, Vancouver, BC, Canada}
\author{D.~Fujimoto}
\affiliation{TRIUMF, Vancouver, BC, Canada}
\author{R.~Fujitani}
\affiliation{Department of Nuclear Engineering, Kyoto University, Kyoto, Japan}
\affiliation{Institute for Integrated Radiation and Nuclear Science (KURNS), Kyoto University, Osaka, Japan}
%\author{D.~Georgescu}
%\affiliation{TRIUMF, Vancouver, BC, Canada}
\author{M.~Gericke}
\affiliation{University of Manitoba, Winnipeg, MB, Canada}
\author{P.~Giampa}
\affiliation{TRIUMF, Vancouver, BC, Canada}
\author{C.~Gibson}
\affiliation{TRIUMF, Vancouver, BC, Canada}
\author{R.~Golub}
\affiliation{North Carolina State University, Raleigh, NC, USA}
\author{K.~Hatanaka}\thanks{deceased}
\affiliation{Research Center for Nuclear Physics (RCNP), The University of Osaka, Osaka, Japan}
\author{T.~Hepworth} \thanks{current affiliation: Physikalisches Institut, Universität Heidelberg, Germany} %ugrad 2023-2025
\affiliation{The University of Winnipeg, Winnipeg, MB, Canada}
\author{T.~Higuchi}
\affiliation{Institute for Integrated Radiation and Nuclear Science (KURNS), Kyoto University, Osaka, Japan}
\affiliation{Research Center for Nuclear Physics (RCNP), The University of Osaka, Osaka, Japan}
%\author{J.~Hussain}
%\affiliation{TRIUMF, Vancouver, BC, Canada}
%\author{G.~Ichikawa}
%\affiliation{High Energy Accelerator Research Organization (KEK), Tsukuba, Ibaraki, Japan}
%\author{I.~Ide}
%\affiliation{Nagoya University, Nagoya, Aichi, Japan}
%\author{S.~Imajo}\thanks{current affiliation:   RIKEN Center for Advanced Photonics, Wako, Saitama, Japan}
%\affiliation{Research Center for Nuclear Physics (RCNP), The University of Osaka, Osaka, Japan}
\author{A.~Jaison}
\affiliation{University of Manitoba, Winnipeg, MB, Canada}
\author{B.~Jamieson}
\affiliation{The University of Winnipeg, Winnipeg, MB, Canada}
\author{K. Jorgensen-Fullam}
\affiliation{TRIUMF, Vancouver, BC, Canada}
\author{M.~Katotoka} % ugrad 2024-2025
\affiliation{The University of Winnipeg, Winnipeg, MB, Canada}
\author{S.~Kawasaki}
\affiliation{High Energy Accelerator Research Organization (KEK), Tsukuba, Ibaraki, Japan}
\affiliation{The Graduate University for Advanced Studies (Sokendai), Tsukuba, Ibaraki, Japan}
\author{M.~Kitaguchi}
\affiliation{Nagoya University, Nagoya, Aichi, Japan}
\author{W.~Klassen}
\affiliation{The University of British Columbia, Vancouver, BC, Canada}
\author{E.~Klemets}
\affiliation{TRIUMF, Vancouver, BC, Canada}
\affiliation{The University of British Columbia, Vancouver, BC, Canada}
\author{E.~Korkmaz}
\affiliation{The University of Northern BC, Prince George, BC, Canada}
\author{E.~Korobkina}
\affiliation{North Carolina State University, Raleigh, NC, USA}
\author{F.~Kuchler}\thanks{current affiliation:  Technical University of Munich, Garching, Germany}
\affiliation{TRIUMF, Vancouver, BC, Canada}
\author{C.~Lamb}
\affiliation{TRIUMF, Vancouver, BC, Canada}
\author{M.~Lavvaf}
\affiliation{University of Manitoba, Winnipeg, MB, Canada}
\author{A.~Lejuez}
\affiliation{TRIUMF, Vancouver, BC, Canada}
\author{T.~Lightbody}
\affiliation{TRIUMF, Vancouver, BC, Canada}
\author{T.~Lindner}
\affiliation{TRIUMF, Vancouver, BC, Canada}
\affiliation{The University of Winnipeg, Winnipeg, MB, Canada}
%\author{N.~Lo} % ugrad fall 2024, where?
%\affiliation{TRIUMF, Vancouver, BC, Canada}
\author{S.~Longo}
\affiliation{University of Manitoba, Winnipeg, MB, Canada}
\author{K.W.~Madison}
\affiliation{The University of British Columbia, Vancouver, BC, Canada}
%\author{Y.~Makida}
%\affiliation{High Energy Accelerator Research Organization (KEK), Tsukuba, Ibaraki, Japan}
%\affiliation{The Graduate University for Advanced Studies (Sokendai), Tsukuba, Ibaraki, Japan}
\author{J.~Malcolm} % ugrad fall 2024, winter 2025
\affiliation{TRIUMF, Vancouver, BC, Canada}
\author{J.~Mammei}
\affiliation{University of Manitoba, Winnipeg, MB, Canada}
\author{R.~Mammei}
\affiliation{The University of Winnipeg, Winnipeg, MB, Canada}
%\author{Z.~Mao}
%\affiliation{The University of British Columbia, Vancouver, BC, Canada}
\author{C.~Marshall}
\affiliation{TRIUMF, Vancouver, BC, Canada}
\author{J.W.~Martin}
\affiliation{The University of Winnipeg, Winnipeg, MB, Canada}
\author{R.~Matsumiya}
\affiliation{TRIUMF, Vancouver, BC, Canada}
\affiliation{Research Center for Nuclear Physics (RCNP), The University of Osaka, Osaka, Japan}
\author{M.~McCrea}
\affiliation{The University of Winnipeg, Winnipeg, MB, Canada}
\author{E.~Miller}
\affiliation{The University of British Columbia, Vancouver, BC, Canada}
\author{M.~Miller}
\affiliation{McGill University, Montreal, QC, Canada}
\author{K.~Mishima}
\affiliation{Research Center for Nuclear Physics (RCNP), The University of Osaka, Osaka, Japan}
\affiliation{Nagoya University, Nagoya, Aichi, Japan}
\affiliation{High Energy Accelerator Research Organization (KEK), Tsukuba, Ibaraki, Japan}
\author{T.~Mohammadi}
\affiliation{University of Manitoba, Winnipeg, MB, Canada}
\author{T.~Momose}
\affiliation{The University of British Columbia, Vancouver, BC, Canada}
\affiliation{TRIUMF, Vancouver, BC, Canada}
\author{M.~Nalbandian} % ugrad 2025
\affiliation{The University of British Columbia, Vancouver, BC, Canada}
\author{T.~Okamura}
\affiliation{High Energy Accelerator Research Organization (KEK), Tsukuba, Ibaraki, Japan}
\affiliation{The Graduate University for Advanced Studies (Sokendai), Tsukuba, Ibaraki, Japan}
%\author{H.J.~Ong}
%\affiliation{Research Center for Nuclear Physics (RCNP), Osaka University, Osaka, Japan}
\author{S.~Pankratz} % ugrad 2025
\affiliation{The University of Winnipeg, Winnipeg, MB, Canada}
\author{R.~Patni} % ugrad fall 2024, winter 2025
\affiliation{TRIUMF, Vancouver, BC, Canada}
\author{R.~Picker}
\email{rpicker@triumf.ca}
\affiliation{TRIUMF, Vancouver, BC, Canada}
\affiliation{Simon Fraser University, Burnaby, BC, Canada}
\author{V.~Purcell}
\affiliation{TRIUMF, Vancouver, BC, Canada}
\author{K.~Qiao}
\affiliation{Research Center for Nuclear Physics (RCNP), The University of Osaka, Osaka, Japan}
\affiliation{Graduate School of Science, The University of Osaka, Osaka, Japan}
%\author{W.D.~Ramsay}
%\affiliation{TRIUMF, Vancouver, BC, Canada}
\author{Y.-N.~Rao}
\affiliation{TRIUMF, Vancouver, BC, Canada}
\author{W.~Rathnakela}
\affiliation{University of Manitoba, Winnipeg, MB, Canada}
\author{T.~Reimer} % ugrad 2025
\affiliation{The University of Winnipeg, Winnipeg, MB, Canada}
\author{D.~Salazar}
\affiliation{Simon Fraser University, Burnaby, BC, Canada}
\affiliation{TRIUMF, Vancouver, BC, Canada}
%\author{A.~Sankaran}
%\affiliation{TRIUMF, Vancouver, BC, Canada}
%\author{J.~Sato}
%\affiliation{Nagoya University, Nagoya, Aichi, Japan}
\author{W.~Schreyer}
%\affiliation{TRIUMF, Vancouver, BC, Canada}
\affiliation{Physics Division, Oak Ridge National Laboratory, Oak Ridge, TN, USA}
\author{T.~Shima}
\affiliation{Research Center for Nuclear Physics (RCNP), The University of Osaka, Osaka, Japan}
\author{H.M.~Shimizu}
\affiliation{Nagoya University, Nagoya, Aichi, Japan}
\author{S. Siddiqui}
\affiliation{TRIUMF, Vancouver, BC, Canada}
\author{S.~Sidhu}
\affiliation{TRIUMF, Vancouver, BC, Canada}
%\author{L.~Smith} % TRIUMF Coop, hasn't worked on source
%\affiliation{The University of British Columbia, Vancouver, BC, Canada}
\author{S.~Stargardter}
\affiliation{University of Manitoba, Winnipeg, MB, Canada}
\author{R.~Stutters} % ugrad 2025
\affiliation{TRIUMF, Vancouver, BC, Canada}
\author{P.~Switzer} % Wpg ugrad summer 2024, summer work and an honours thesis on the UCN guide-coating facility
\affiliation{The University of Winnipeg, Winnipeg, MB, Canada}
%\author{I.~Tanihata}
%\affiliation{Research Center for Nuclear Physics (RCNP), The University of Osaka, Osaka, Japan}
\author{Tushar}
\affiliation{University of Manitoba, Winnipeg, MB, Canada}
%\author{B.~van der Veek}  % MSL person
%\affiliation{???}
\author{M.~Uzair}
\affiliation{TRIUMF, Vancouver, BC, Canada}
\author{S.~Vanbergen}
\affiliation{The University of British Columbia, Vancouver, BC, Canada}
\affiliation{TRIUMF, Vancouver, BC, Canada}
\author{W.T.H.~van~Oers}
\affiliation{TRIUMF, Vancouver, BC, Canada}
%\author{Y.~Watanabe}
%\affiliation{High Energy Accelerator Research Organization (KEK), Tsukuba, Ibaraki, Japan}
\author{N.~Yazdandoost}
\affiliation{TRIUMF, Vancouver, BC, Canada}
\author{Q.~Ye}
\affiliation{The University of British Columbia, Vancouver, BC, Canada}
\author{A.~Zahra}
\affiliation{University of Manitoba, Winnipeg, MB, Canada}
\author{L.~Zhang}  % beam physicist
\affiliation{TRIUMF, Vancouver, BC, Canada}
%\author{M.~Zhao} % ugrad winter-summer 2024 only worked on EDM
%\affiliation{TRIUMF, Vancouver, BC, Canada}

\collaboration{TUCAN Collaboration}%\noaffiliation

\date{\today}% It is always \today, today,
             %  but any date may be explicitly specified

\begin{abstract}
The TRIUMF UltraCold Advanced Neutron (TUCAN) collaboration has completed a new superthermal source for ultracold neutrons (UCNs) at TRIUMF.
Neutrons produced in a spallation target %driven by TRIUMF's %520-MeV main cyclotron.
are moderated in heavy water and liquid deuterium, and are converted to ultracold neutrons by inelastic scattering inside superfluid $^4$He at around \qty{1.1}{\kelvin}.
During commissioning of the completed moderation and conversion system, including the liquid-deuterium moderator, up to $1.47(2)\times 10^7$ UCNs were detected outside the source in the experimental area following a \qty{60}{\second} accumulation period.
%During continuous operation, a rate of up to \qty{6.75(3)e5}~UCN/s was detected in the same location outside the radiation shielding.
%A density of \qty{165.6(1)}{UCN \per \centi \cubic \meter} was achieved in a \qty{25.4}{\liter} vessel in the experimental area.
At the same location, continuous operation produced detector rates of up to \qty{6.75(3)e5}{UCN/s}, and a density of \qty{165.6(1)}{UCN \per \centi \cubic \meter} was achieved in a \qty{25.4}{\liter} external vessel.
These measurements demonstrate high-intensity operation of a spallation-driven He-II UCN source, including extraction and transport to an external experimental area.

%This is the world's highest published UCN rate outside the source.
%This corresponds to a density of
%Up to \qty{6.75(3)e5} UCN/s were detected during continuous operation, more than at any other source in the world.
\end{abstract}

%Use showkeys class option if keyword
                              %display desired
\maketitle

%TC:endignore
% \textcolor{purple}{From here to acknowledgements: 3750 words and $<$ 4 pages: \url{https://journals.aps.org/authors/length-guide}.} 
%\\
%%%%%%%%%%%%%%%%%%%%%%%%%%%%%%%%%%%%%%%%%%%%%%%%%%%%%%%%
%\tableofcontents
{\it Introduction}---
Ultracold neutrons (UCNs) are free, unbound neutrons with extremely low kinetic energies (typically below \qty{300}{\nano \electronvolt}) and corresponding velocities $<$~\qty{8}{\meter \per \second}.
Their defining property is that they reflect off certain material surfaces at any angle of incidence, allowing them to be physically trapped in ``bottles'' %or magnetic arrays
for hundreds of seconds.
This long observation time allows exceptionally precise measurements of the neutron's fundamental properties,
such as the neutron lifetime~\cite{serebrov2005,musedinovic2025}, its electric dipole moment (EDM)~\cite{ILLEDMPendl2015,bib:psi2020,bib:panedm,bib:n2edm,bib:ss}, $\beta$-decay correlation parameters~\cite{brown2018new} and investigations of quantum states of
neutrons in the Earth's gravitational
field~\cite{nesvizhevsky2002quantum,cronenberg2018acoustic}.
Together, they provide fundamental tests of the Standard Model of particle physics and constraints on cosmological models.
%Together, they provide fundamental tests of the Standard Model of particle physics and cosmology.

These experiments are, in many cases, limited
by statistical precision, mainly due to the low achievable phase-space
density. % and finite storage lifetimes of UCNs. 
Statistical uncertainties scale inversely with square root of the number of detected
UCNs. % and their
%survival probability during storage, transport, and measurement. 
As a result, increased UCN production rates and reduced losses are key
drivers for improved sensitivity across these experiments, as well as improved studies of
systematic effects.

For decades, the turbine UCN source at the Institut Laue-Langevin (ILL) provided the highest UCN densities to experiments~\cite{crow1995PF2}.
It Doppler-shifts neutrons from the cold source of the research reactor~\cite{AGERON1989197ILLColdSources} by reflecting them from rapidly receding mirrors mounted to a spinning turbine.
The neutrons lose a significant portion of their kinetic energy and become ultracold.
An alternative approach is superthermal UCN production~\cite{GOLUB1975133}.
In a superthermal source, incident neutrons are down-scattered into the UCN energy range by creating collective excitations in a cryogenic converter.
The reverse process is suppressed when the converter is maintained at sufficiently low temperature because the thermal population of the relevant excitations is small.
This principle has been implemented using either solid deuterium or superfluid helium. 

Several superthermal UCN sources are currently operational:
The Paul Scherrer Institute (PSI) and the Los Alamos National Laboratory (LANL) employ spallation-driven neutron sources with solid ortho-deuterium (SD$_2$) as the converter~\cite{bib:lanlsource,bib:psisource}.
At Johannes Gutenberg University Mainz, solid-deuterium UCN sources can be operated with the TRIGA reactor in continuous or pulsed mode~\cite{karch2014mainzperf,kahlenberg2017upgrade}.
The SuperSUN project utilizes a superfluid $^4$He (He-II) converter in a cold-neutron beamline at the continuously operating ILL research reactor~\cite{bib:supersun}.
The TRIUMF UltraCold Advanced Neutron (TUCAN) source, the subject of this paper, employs a He-II converter directly coupled to a spallation target.
Although the UCN production cross section is larger in
SD$_2$~\cite{bib:yumalikgolub,bib:atchison,bib:frei} than in
He-II~\cite{bib:golub77}, upscattering and material effects in SD$_2$ limit UCN lifetimes to the millisecond timescale~\cite{bib:lanlucn,bib:chenyu,bib:psincsulanl}.
In He-II, losses are dominated by phonon upscattering and predominantly scale with temperature as
$T^7$~\cite{bib:zimmer,bib:leung,bib:yoshiki};  below
\SI{1}{\K}, storage lifetimes exceeding \SI{100}{\s} have been
demonstrated~\cite{bib:yoshiki}.

The TUCAN He-II source utilizes a proton beam to drive its dedicated spallation target \cite{vanbergenintds2023}. %, which is coupled to a room-temperature nEDM apparatus.
% utilizes a proton beam to drive a spallation-based He-II UCN source coupled to a room-temperature nEDM apparatus. 
H$^-$ ions are injected and accelerated in the TRIUMF cyclotron; a thin carbon foil strips their two electrons, and 483~MeV protons are extracted into beamline 1. % (BL1).
A fast kicker magnet~\cite{bib:kicker} diverts a fraction of the proton pulses onto the tungsten neutron-spallation target in beamline 1U~\cite{bib:ucnbeamline}.
A heavy-water (D$_2$O)  and (for the first time) directly coupled liquid-deuterium (LD$_2$) moderator increase the cold neutron flux inside the volume where the UCNs are produced.
We have previously reported results from the partially completed TUCAN source without the LD$_2$ moderator~\cite{TUCANinitialproduction}.
%This letter contains results of source characterization with a filled LD$_2$ moderator vessel, and with detection moved to the experimental area in the Meson Hall, outside the radiation shielding.
The present work reports the first characterization of the complete moderation and conversion chain, including the filled liquid-deuterium moderator, accumulation in the He-II converter, and transport through the radiation shielding to the external experimental area.
We characterize the source in batch and continuous operating modes and by storing UCN in the experimental area.

%%%%%%%%%%%%%%%%%%%%%%%%%%%%%%%%%%%%%%%%%%%%%%%%%%%%%%%%
%TC:ignore
\begin{figure}[htb] 
%trim=left bottom right top
\centering \includegraphics[width=\columnwidth, page=2,trim=4.3cm 2.7cm 0cm 3.1cm,clip]{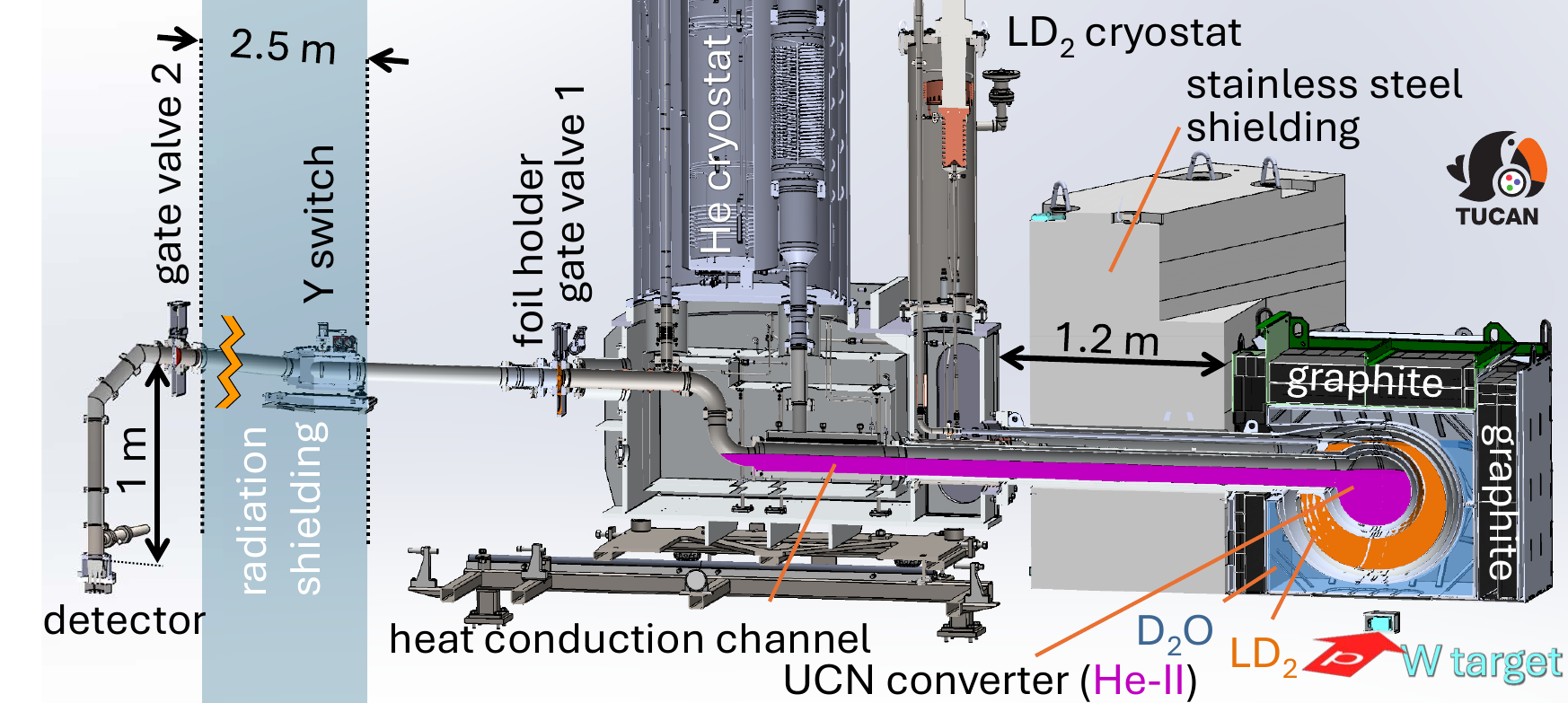}
\caption{Cut-open view of the TUCAN source and guides to the East port. A red arrow indicates the proton beam impinging on the tungsten (W) target.
The approximate fill levels of UCN moderator and converter liquids during experiments are indicated.
The UCN guide in the figure is shortened left of the Y switch as indicated by the black-orange jagged line.\label{fig:setup}}
\end{figure}
%TC:endignore
{\it The experimental setup}---
%The TRIUMF cyclotron provides protons with an energy of \SI{483}{\mega \electronvolt} to BL1U that contains the water-cooled tungsten (W) target~\cite{bib:ucnbeamline}, where spallation neutrons are generated.
Free neutrons created by spallation in the water-cooled tungsten target~\cite{bib:ucnbeamline} first traverse a 1-inch-thick layer of lead that serves as a radiation shield~\cite{bib:moderators}.
D$_2$O at room temperature and LD$_2$ at \qty{25(1)}{\kelvin} moderate neutrons %towards %the thermal and then cold neutron regions, respectively, 
as shown in Fig.~\ref{fig:setup}. 
Both are held in aluminum vessels.
A graphite layer, approximately \qty{10}{\centi \meter} thick, surrounds the moderator region and acts as a neutron reflector.
The central volume contains the He-II at \qty{1.1(2)}{\kelvin}, which converts cold neutrons to ultracold neutrons.
%Its aluminum vessel is coated with nickel-phosphorus (NiP, 10-13~wt\% P content) using electroless nickel plating
Its aluminum vessel is coated by electroless plating with nickel-phosphorus (NiP) containing 10–13\% phosphorus by mass~\cite{akatsuka2023characterization}.
UCNs are internally reflected from this coating and transported through the liquid helium.
%to the experiments at room temperature.
The He-II filled channel also serves to transport heat from the production volume to the He cryostat that provides up to \qty{10}{\watt} of cooling power~\cite{kawasaki2019cryogenic}.
Gate valve 1 (see Fig.~\ref{fig:setup}), right outside the cryostat, allows storage of UCNs inside the source.
Due to limited stock during the experiments described herein, the heavy water moderator vessel was filled to \SI{410(20)}{\liter}, corresponding to 72(4)\% of the full volume. % 573 l new info from Sean
The near-spherical UCN production volume contained He-II to a fill level of 27(1) cm out of a total diameter of 36 cm (85(5)\% of the production volume).
The horizontal heat conduction channel contained liquid up to 50(5)\% of its full volume.
In contrast to Ref.~\cite{TUCANinitialproduction}, the cold moderator system was operational and fully filled with LD$_2$. % shown above already: at a temperature of \qty{25(1)}{K}.

Directly downstream of gate valve 1, a titanium foil with a thickness of \qty{15}{\micro \meter} was installed during some experiments to prevent contamination of the cryogenic UCN production volume;
a rotary Y switch directed UCNs to either the East port (EDM experiment) or the West port (second experimental area). No distinction between the two ports is made below, as their results were statistically indistinguishable. UCN guide sections connected the source to both beam ports through the 2.5-m-thick radiation shielding.
Just outside of the shielding, a 90-degree bend downward directly after gate valve 2 (see Fig.~\ref{fig:setup}) provided enough gravitational acceleration with a drop of \qty{1.02}{m} %1068.4mm-48.2mm drop from bottom of horizontal guide to Li surface
for the UCNs to overcome the Fermi potential of the lithium-6-loaded scintillating glass detector (\qty{103(1)}{\nano \electronvolt})~\cite{ban2016lithium}. 

Two experimental campaigns, denoted A and B, were conducted in December 2025 and January 2026, respectively.
A vacuum leak in the guide system at the beginning of campaign A during a period without a Ti foil allowed air ingress and contamination of the cryogenic guides, suppressing the UCN yield.
Source performance was recovered before campaign B by warming and baking the source.

%%%%%%%%%%%%%%%%%%%%%%%%%%%%%%%%%%%%%%%%%%%%%%%%%%%%%%%%
{\it UCN production and storage in the source}---
We used a very similar UCN production cycle as in previous experiments~\cite{bib:physrevc,TUCANinitialproduction}, which we denote batch production:
two gate valves as shown in Fig.~\ref{fig:setup} were closed while the spallation target was irradiated with proton currents between 1 and \qty{33}{\micro \ampere}.
The design current of \qty{40}{\micro \ampere} could not be reached for most of the experiments due to limitations of the current control firmware of the kicker magnet.
An improved version is scheduled to be implemented in the next year.
% \textcolor{red}{Make clearer that we can easily fix this.}
After the beam was turned off, the gate valves would open after a delay between 0 to \qty{160}{\second}.
During this period, UCNs are stored in the source.
Details about the uncertainty determination and background subtraction for the data in this report are described in the Appendix.% ~\ref{app:uncertainty}. 

%TC:ignore
\begin{figure}[ht]
\begin{center}
  \includegraphics[width=\columnwidth]
  {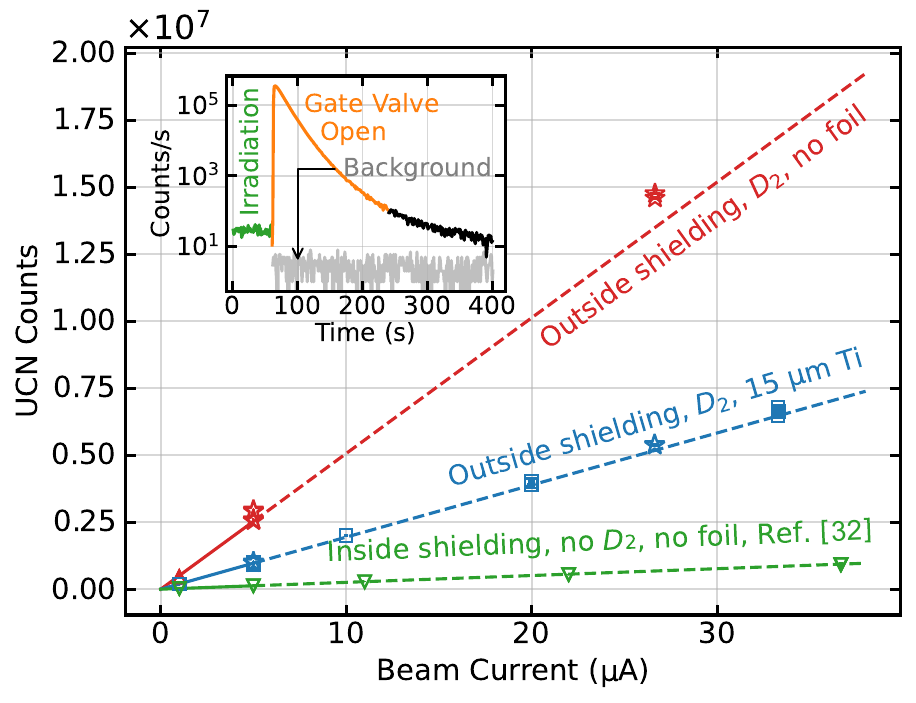}
\caption{
UCN counts integrated over the \SI{180}{\s} counting period as a function of the average beam current during the preceding \SI{60}{\s} irradiation period, for three different configurations. Both gate valves were opened immediately after the irradiation period. 
\textcolor{darkgreen}{$\pmb{\triangledown}$}:~Ref.~\cite{TUCANinitialproduction}, 
\textcolor{red}{$\pmb{\vartriangle}$}\textcolor{RoyalBlue}{\scalebox{0.85}{$\pmb{\square}$}}: campaign A -  
\thickredstar\thickbluestar:~campaign B.
UCN counts are greatly increased by the LD$_2$ moderator.  
The data \qty{\leq 5}{\micro \ampere} are fitted to a linear model with zero intercept (solid lines) and are extrapolated (dashed) to the highest currents. 
%Deviation of the integrated counts from the trend line is small and mostly visible for the data points from campaign B (\thickredstar,\thickbluestar), which trend higher. 
Errors are drawn and often smaller than marker sizes. Inset: Raw counts from a typical measurement cycle at \SI{33.3(3)}{\uA} with foil. Backgrounds were measured by keeping the gate valve closed throughout the entire cycle. In this case, we find $487(16)$ background counts (grey) and $6.795(3)\times 10^6$ total counts (orange) with the valve open.
}
\label{fig:batch}
\end{center}
\end{figure}
%TC:endignore
Example experimental cycles with zero delay shown in the inset of Fig.~\ref{fig:batch} exhibit
significantly larger count rates with the gate valves open (orange) with the deuterium moderator compared to Ref.~\cite{TUCANinitialproduction} without it.
Additionally, the count rates during target irradiation (green) are about three orders of magnitude lower since the detector is outside of the radiation shielding.
To determine electronic and detector background, we kept the two gate valves closed after the irradiation period (grey). 
We repeated the experiment with and without a \qty{15}{\micro \meter} Ti foil placed inside the UCN guide
after gate valve 1 and for varying proton currents, as shown in the main plot of Fig.~\ref{fig:batch}.
With higher beam currents, the UCN production increases, but the production volume also receives more heat from the spallation target, leading to higher He-II temperatures.
Monte Carlo simulations~\cite{bib:moderators} predict a UCN production rate of \qty{2.76e5}{\per \second \per \micro \ampere} and a heat load of \qty{0.167}{\watt \per \micro \ampere}.
Since UCN upscattering scales as $T^7$, shorter storage lifetimes of UCNs in the source are expected.
Therefore, we anticipated the dependence of the UCN counts as a function of beam current to deviate from linearity at higher currents.
However, as reported in~\cite{TUCANinitialproduction}, a nearly linear correlation is observed:
trend lines in Fig.~\ref{fig:batch} using only the data points up to \qty{5}{\micro \ampere} and extrapolated to higher currents indicate basically no deviation at high currents.
%The highest number of neutrons measured outside of the radiation shielding for batch production was % (old) 13453136.5456 ± 134622.4338, (old) 13369572.12019858 +- 133745.7213 
% (new) 14728010.754297117 ± 147330.0991483703
We detected up to $1.47(2)\times 10^7$ UCN during campaign B for a current of %26.63816 ± 0.26638 
\qty{26.6(3)}{\micro \ampere}. 
This is only surpassed by the PSI UCN source~\cite{bib:psisource}.
With a source volume of \qty{88.6}{\liter} up to gate valve 1, this amounts to a density of \qty{166(2)}{\per \centi\cubic\meter} %\qty{151(2)}{\per \centi\cubic\meter}
in the source that can be extracted and detected.

A comparison of the linear fit constants shows that the UCN counts increased by a factor of $19.9(2)$ % was $20.0(2)$ then $19.2(2)$
between production runs \textsl{without} deuterium, where UCN are detected \textsl{inside} the radiation shielding (green data points, from~\cite{TUCANinitialproduction}), and production runs \textsl{with} deuterium, where UCN are detected \textsl{outside} the radiation shielding and the path length increases by around \qty{2.5}{\meter} (red data points).
Adding the Ti foil reduced the UCN counts to 38.4(2)\% (red\,$\rightarrow$\,blue). 

%TC:ignore
\begin{figure}[ht]
\begin{center} 
%trim=left bottom right top
  \includegraphics[width=\columnwidth, page=1,trim=0cm 0cm 0cm 0cm,clip]{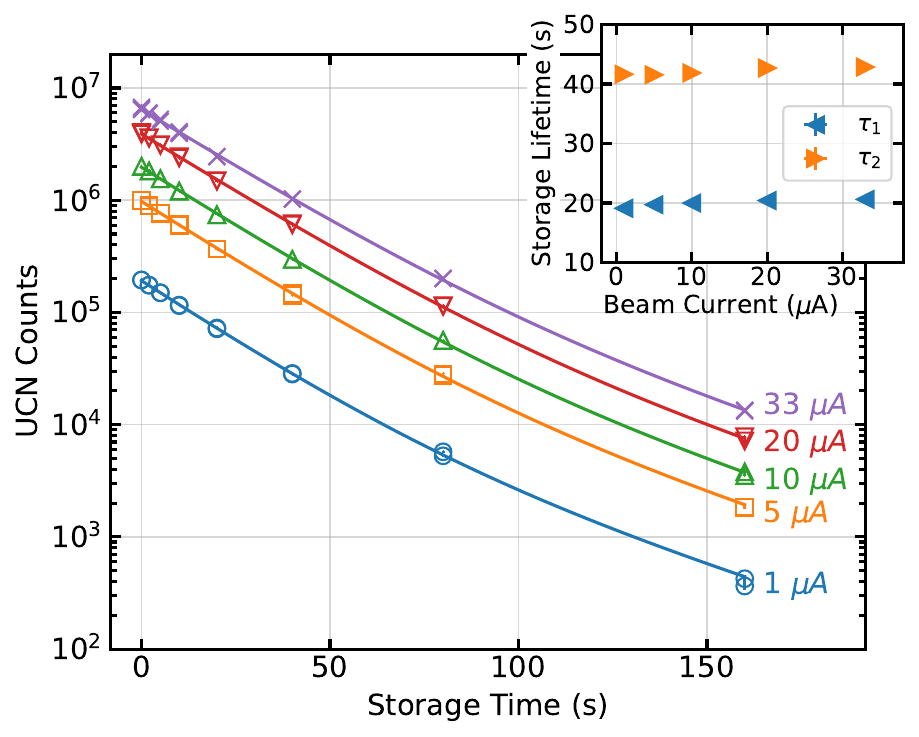}
\caption{
%Main plot: UCN counts in the detector as a function of storage time in the TUCAN source for different proton currents. Inset: Storage lifetimes as a function of proton current.} %
Detected UCNs after \SI{60}{\s} irradiation, for varying storage times during which the beam is off. 
Bi-exponential decays are fit to the counts.
Uncertainties are drawn and small compared to marker sizes.
Inset: Storage lifetime constants as a function of proton current determined from bi-exponential fits show a slightly increasing trend for higher currents. %The exponential lifetime increases by \SI{0.5}{\s} for currents of \SI{20}{\uA} and higher. 
Fit parameters are shown in Table~\ref{tab:table1}.
}
\label{fig:SSL}
\end{center}
\end{figure}
%TC:endignore
To perform \emph{UCN storage-lifetime experiments inside the source volume}, the delay in opening the gate valves after irradiation was varied.
Results are shown in Fig.~\ref{fig:SSL} for different beam currents.
We fit the data with a bi-exponential function $N(t)=N_0\left(fe^{-t/\tau_1}+(1-f)e^{-t/\tau_2}\right)$ over the whole range;
the average results are $\tau_1 =$ \qty{20.0(5)}{\second} and $\tau_2 =$ \qty{42.1(5)}{\second}.
Higher currents cause increased radiation heating of the UCN source, leading to higher temperatures of the superfluid helium.
This increases the UCN upscattering and should lead to shorter storage lifetimes.
Instead, the time constants trend longer for higher currents.
We are planning to perform dedicated experiments to investigate this further.
%Storage lifetimes of UCNs in material vessels are energy dependent~\cite{Gol91}.
%The energy spread in the TUCAN source can be quite large (up to \qty{195}{\nano \electronvolt}), determined by the difference between the Fermi potential of superfluid helium (\qty{18.5(2)}{\nano \electronvolt}~\cite{sears1992neutron}) and the NiP coating of the source volumes (\qty{213(5)}{\nano \electronvolt}~\cite{PATTIE2017NiP}). % Error propagation here: https://share.google/aimode/Xa85iV6CytKQdX64o
%Hence, the reduced $\chi^2$ values are rather large; see Table~\ref{tab:table1} in the Appendix.%~\ref{app:fitparams}. 

%TC:ignore
\begin{figure}[htb]
%trim=left bottom right top
\centering \includegraphics[width=\columnwidth, page=1,trim=0cm 0.3cm 0cm 0cm,clip]{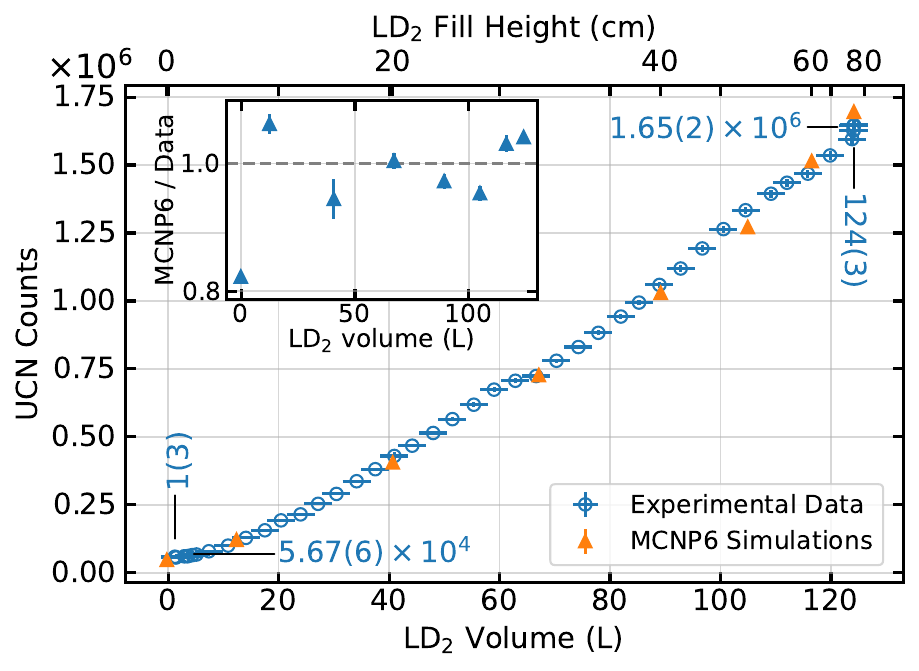}
\caption{Ultracold neutrons detected outside the radiation shielding as a function of LD$_2$ volume in the moderator vessel. 
Predictions of the UCN production change from MCNP6~\cite{MCNP61} are also shown.
The counts are integrated over a \qty{340}{\second} period and normalized to \SI{5}{\micro \ampere} by averaging the proton current over the irradiation period (\qty{60}{\second}).
The UCN yield increases by a factor of 29.1(5) overall. 
Simulation and experimental results are normalized using their mean values for data points where uncertainties overlap with respect to LD$_2$ volume.
Uncertainties are displayed but mostly smaller than the marker symbols.
Inset: Ratio of simulation and data showing the maximum deviation for an empty LD$_2$ vessel.
\label{fig:batchvsD2level} }
\end{figure}
%TC:endignore
We determined the \emph{effect of the cold moderator on UCN production} by conducting repeated batch production cycles while recovering the liquid deuterium from its vessel. % as shown in Fig.~\ref{fig:batchvsD2level}.
To compare the results with expectations, we performed simulations using Monte Carlo N-Particle$^{\circledR}$ code MCNP6~\cite{MCNP61}.
 %compare the results with expectations, we performed simulations using Monte Carlo N-Particle$^{\circledR}$ code MCNP6~\cite{MCNP61}.
In the simulation, \num{483}\text{-}\si{\mega \electronvolt} protons hit the spallation target, and the secondary particles are tracked in the as-built geometry and materials of the moderators and UCN source~\cite{bib:moderators}.
A neutron scattering kernel specifically designed for He-II is employed~\cite{lavelleHeKernel}.
UCN production in the converter volume %, see Fig.~\ref{fig:setup}, 
is calculated from the differential neutron flux multiplied by the energy-dependent UCN production cross section from~\cite{bib:leung,bib:schmidt-wellenburg}.
Fig.~\ref{fig:batchvsD2level} shows the UCN count data as a function of LD$_2$ fill volume.
The inset shows that agreement is generally within 5\% except for empty LD$_2$ vessel.
The UCN yield from the TUCAN source increased by a factor of 29.1(5) from an empty LD$_2$ moderator to a full one.
This is higher than deduced from Fig.~\ref{fig:batch}, since the guide layout remained unchanged.

%%%%%%%%%%%%%%%%%%%%%%%%%%%%%%%%%%%%%%%%%%%%%%%%%%%%%%%%
%TC:ignore
\begin{figure}[htb]
%trim=left bottom right top
\centering \includegraphics[width=\columnwidth, page=1,trim=0cm 0cm 0cm 0cm,clip]{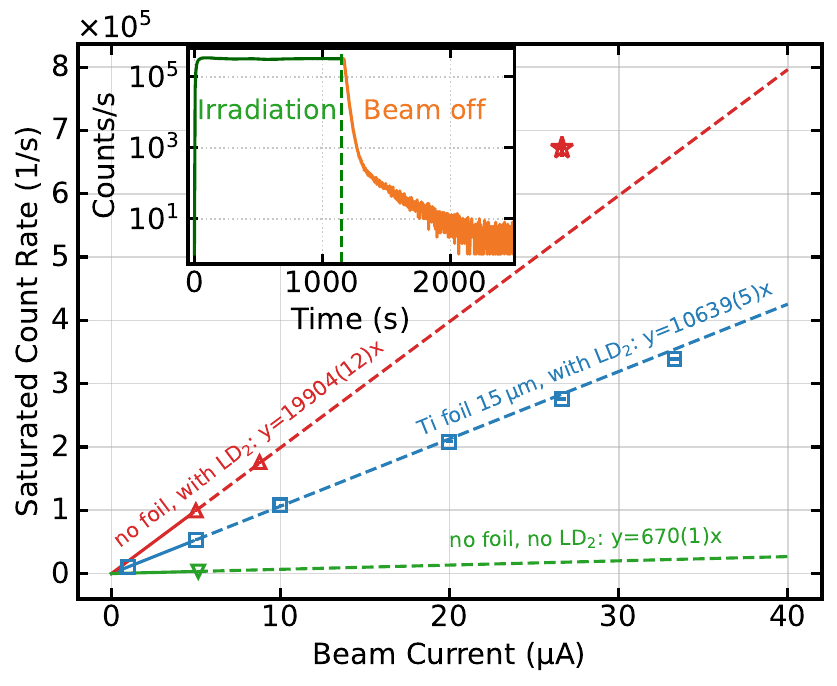}
\caption{Saturated ultracold neutron count rates detected outside the radiation shielding during continuous UCN production as a function of BL1U beam current. Uncertainties in beam current and count rate are displayed but generally smaller than the marker symbols. 
\textcolor{darkgreen}{$\pmb{\triangledown}$}\textcolor{RoyalBlue}{\scalebox{0.85}{$\pmb{\square}$}}\textcolor{red}{$\pmb{\vartriangle}$}: campaign A -  
\thickredstar:~campaign B.
%All data from campaign A, except \thickredstar~from campaign B. 
Inset: UCN counts as a function of time with \qty{33}{\micro \ampere} target irradiation and a Ti foil. The beam-on period (\qty{1150}{\second}, green) shows that the count rate stays largely constant during irradiation once saturated. The beam-off period is plotted in orange.}
\label{fig:SSP} 
\end{figure}
%TC:endignore
{\it Continuous or steady-state UCN production---} 
The TUCAN source also has the ability to continuously produce UCNs, which is very beneficial for filling large experiments or for flow-through experiments.
In this regime, the target is irradiated while all gate valves are open.
The inset of Fig.~\ref{fig:SSP} shows the count rate in the detector as a function of time. %: it shows saturation behavior during the irradiation period. % and drops with a bi-exponential function when the beam is turned off.
We fitted exponential saturation curves to the first \qty{200}{\second} of the irradiation periods to determine saturated count rates.
The beam-off period exhibits two exponential decay constants plus a constant background.
We associate the fast decay with UCN draining to the detector and the slow one with activation of the aluminum detector housing by UCN.
The activation and constant background were subtracted as laid out in the appendix.
The background-corrected saturated count rates are displayed in the main plot as a function of beam current.

For each configuration, with and without foil and LD$_2$, linear fits up to \qty{5}{\micro \ampere} were performed and extrapolated to \qty{40}{\micro \ampere}.
The count rates for the configuration with foil and LD$_2$ show no significant deviation from linearity towards higher currents, similar to batch production results. % and even higher yields are possible.
%The inset shows one example at \qty{33}{\micro \ampere}, where it can be seen that the count rate stays largely constant for more than 1000~s.
%The data set without foil was not complete enough to perform a similar linear fit. \textcolor{red}{try fit with just one data point}
All data points in Fig.~\ref{fig:SSP} belong to campaign A, with the exception of the highest point, which deviates above the extrapolated linear fit:
during campaign B, up to \qty{6.75(3)e5} UCN/s were detected at a proton current of \qty{26.6(3)}{\micro \ampere} (with \(\text{LD}_{2}\), without Ti foil).
This rate detected \emph{outside the source} exceeds any other UCN facility~\cite{bib:supersun,SEREBROV2009263,DOGE2020163112,bib:lanlsource,kahlenberg2017upgrade}. %
Steyerl et al. reported a higher count rate at PF2 ILL in 1986~\cite{STEYERL1986347}.
However it was extrapolated to the surface of the ILL turbine \emph{inside the vacuum vessel} of the source and cannot be compared to an extracted, detected rate.

{\it Storage outside the shielding---} 
During campaign B, we also delivered (unpolarized) UCNs to a prototype EDM cell in the experimental area, contained them using a valve and then emptied them into a detector.
The cylindrical storage vessel  (height \qty{0.12}{\meter}, inner diameter \qty{0.5}{\meter},  volume \qty{25.4}{\liter}) is coated with NiP~\cite{Vanbergen_PhD_2025}.
It was placed at the elevation of the gate valves (see Fig.~\ref{fig:setup}) and connected via roughly 7~m of UCN guides downstream of gate valve 1.
During our only period at nominal proton current of \qty{39.6(1)}{\micro \ampere}, \qty{4.207(3)e6} and 
\qty{7.21(1)e5} UCN were detected without foil for \qty{0}{\second} and \qty{170(1)}{\second} storage time, respectively, and after background subtraction.
This corresponds to directly-detectable densities of %165.6261399	0.1159449506
$165.6(1)$ and %28.39066398	0.03615187231
\qty{28.40(4)}{UCN \per \centi \cubic \meter}, respectively, an unprecedented UCN density stored in a vessel outside a UCN source~\cite{STEYERL1986347, DOGE2020163112,ito2018performance,kahlenberg2017upgrade, ries2016comparison, WONG2023168105}.
No correction for collection or detection efficiency was made.
More details will be described in a dedicated publication.

%%%%%%%%%%%%%%%%%%%%%%%%%%%%%%%%%%%%%%%%%%%%%%%%%%%%%%%%
{\it Summary and outlook}---
We demonstrate high-intensity operation of the complete TUCAN spallation-driven superfluid-helium UCN source, including heavy-water and liquid-deuterium moderation, extraction, and delivery to an external experimental area. The measurements establish the performance of the full source architecture in both batch and continuous modes and with respect to UCN storage in the experimental area.

Up to %\qty{1.34(1)e7} this value is before the correction
$1.47(2)\times 10^7$ neutrons were detected after irradiating the spallation target with a proton current of \qty{26.6(3)}{\micro \ampere} for 60~s while keeping the UCN valve closed and then immediately opening it to the UCN detector (batch production).
%Compared to the first UCN production with emtpy LD$_2$ vessel~\cite{TUCANinitialproduction}, the UCN counts increased by a factor of 28.73(28), around 20\% less than predicted by Monte Carlo simulations.
The UCN yield increases by a factor of 29.1(5) %28.73(28)
with LD$_2$ moderator. %while Monte Carlo simulations predicted a factor of 35.3(3).
UCN count rates as high as \qty{6.75(3)e5} UCN/s were demonstrated in the experimental area for continuous UCN production with UCN valves open.
For both batch and continuous UCN production, yields increase nearly linearly with higher proton currents.
Additionally, we stored a density up to 
\qty{165.6(1)}{UCN \per \centi \cubic \meter} in an external storage volume of \qty{25.4}{\liter}.
%UCN counts of up to \qty{2e6} for batch production and \qty{1e6} UCN/s are achievable once we begin operating the TUCAN source at the nominal current of \SI{40}{\uA}.
%Filling the heavy water moderator and the UCN production vessel completely is expected to lead to further incremental increases of the UCN production. % by around xx\%.
This opens up new possibilities to improve statistical sensitivities of many ultracold neutron experiments.

%TC:ignore
%\\ \textcolor{purple}{Anything below does not count towards page or word limit.} \\
% ==========================================
% 1. CORE TEXT ENDS HERE
% ==========================================
{\it Data availability}---
The data supporting the findings of this study can be made available by the corresponding author upon reasonable request.

{\it Acknowledgments}---
%\begin{acknowledgments}
We would like to sincerely thank the Mechanical Engineering Center at KEK and the TRIUMF Design and Fabrication, Cryogenic, Accelerator Systems, and Operations groups and
acknowledge the important conceptual contributions by Y.~Masuda.
We gratefully acknowledge the support of the Canada Foundation for Innovation (CFI);
the Canada Research Chairs program;
the Natural Sciences
and Engineering Research Council of Canada (NSERC) SAPPJ-2016-00024,
SAPPJ-2019-00031, SAPPJ-2023-00029, and SAPPJ-2024-00030;
British Columbia Knowledge Development Fund;
Research Manitoba Grant RMB NIOG A\#7213; 
the BC DRI Group and the Digital Research Alliance of Canada;
JSPS KAKENHI (Grant Nos. 18H05230, 19K23442, 20KK0069, 20K14487, and 22H01236, 25H00652);
JSPS Bilateral Program (Grant No. JSPSBP120239940);
JST FOREST Program
(Grant No.  JPMJFR2237);
International Joint Research Promotion
Program, Global Expansion Research Program and Global Academic Collaboration Program of the University of Osaka; RCNP COREnet;
the Yamada Science Foundation; 
the Murata Science Foundation;
the Grant for Overseas Research by the Division of Graduate Studies (DoGS) of Kyoto University; 
the Universidad Nacional Aut\'onoma de M\'exico -
DGAPA program PASPA and grant PAPIIT AG102023;
and the U.S. Department of Energy under contract numbers DE-AC05-00OR22725, KB0401072 and 560233.
%\end{acknowledgments}

{\it Author Contributions}---
\textbf{K.~Abe:} investigation (Minor (Supporting)). \textbf{S.~Ahmed:} investigation (Minor (Supporting)); resources (Minor (Supporting)). \textbf{B.~Algohi:} resources (Minor (Supporting)). \textbf{D.~Anthony:} resources (Minor (Supporting)). \textbf{L.~Barr\'on-Palos:} funding acquisition (Major (Equal)); investigation (Minor (Supporting)); writing – review \& editing (Major (Equal)). \textbf{Y.~Bylinsky:} project administration (Minor (Supporting)); supervision (Minor (Supporting)); validation (Minor (Supporting)). \textbf{M.~Boss\'e:} investigation (Minor (Supporting)); resources (Minor (Supporting)). \textbf{M.P.~Bradley:} funding acquisition (Major (Equal)). \textbf{A.~Brossard:} investigation (Lead); project administration (Minor (Supporting)); resources (Major (Equal)); supervision (Minor (Supporting)); writing – review \& editing (Major (Equal)). \textbf{J.~Chak:} project administration (Major (Equal)); resources (Major (Equal)). \textbf{R.~Chiba:} investigation (Minor (Supporting)); resources (Minor (Supporting)). \textbf{C.~Davis:} conceptualization (Minor (Supporting)); resources (Minor (Supporting)); writing – review \& editing (Minor (Supporting)). \textbf{R.~de~Vries:} resources (Minor (Supporting)). \textbf{K.~Dong:} formal analysis (Lead); investigation (Minor (Supporting)); resources (Minor (Supporting)); writing – original draft (Minor (Supporting)). \textbf{K.~Drury:} formal analysis (Minor (Supporting)); resources (Minor (Supporting)). \textbf{B.~Franke:} funding acquisition (Major (Equal)). \textbf{D.~Fujimoto:} data curation (Lead); formal analysis (Major (Equal)); investigation (Major (Equal)); software (Lead); writing – review \& editing (Major (Equal)). \textbf{R.~Fujitani:} investigation (Minor (Supporting)). \textbf{M.~Gericke:} funding acquisition (Major (Equal)). \textbf{P.~Giampa:} funding acquisition (Minor (Supporting)); investigation (Minor (Supporting)); resources (Minor (Supporting)); software (Minor (Supporting)). \textbf{C.~Gibson:} investigation (Minor (Supporting)); project administration (Lead). \textbf{R.~Golub:} conceptualization (Minor (Supporting)); funding acquisition (Major (Equal)); writing – review \& editing (Major (Equal)). \textbf{K.~Hatanaka:} conceptualization (Major (Equal)); funding acquisition (Major (Equal)); methodology (Major (Equal)); project administration (Major (Equal)); resources (Major (Equal)); supervision (Major (Equal)). \textbf{T.~Hepworth:} investigation (Minor (Supporting)); resources (Minor (Supporting)). \textbf{T.~Higuchi:} conceptualization (Minor (Supporting)); funding acquisition (Major (Equal)); investigation (Minor (Supporting)); methodology (Minor (Supporting)); resources (Major (Equal)); writing – review \& editing (Major (Equal)). \textbf{A.~Jaison:} investigation (Minor (Supporting)). \textbf{B.~Jamieson:} resources (Minor (Supporting)). \textbf{Karina~Jorgensen-Fullam:} resources (Minor (Supporting)). \textbf{M.~Katotoka:} investigation (Minor (Supporting)); resources (Minor (Supporting)). \textbf{M.~Kitaguchi:} funding acquisition (Major (Equal)); supervision (Minor (Supporting)). \textbf{W.~Klassen:} investigation (Minor (Supporting)). \textbf{E.~Klemets:} investigation (Minor (Supporting)); writing – review \& editing (Minor (Supporting)). \textbf{E.~Korkmaz:} funding acquisition (Major (Equal)). \textbf{E.~Korobkina:} funding acquisition (Major (Equal)); writing – review \& editing (Major (Equal)). \textbf{F.~Kuchler:} conceptualization (Major (Equal)); resources (Major (Equal)). \textbf{M.~Lavvaf:} investigation (Major (Equal)). \textbf{C.~Lamb:} resources (Minor (Supporting)). \textbf{A.~Lejuez:} formal analysis (Minor (Supporting)); writing – review \& editing (Major (Equal)). \textbf{T.~Lightbody:} resources (Minor (Supporting)). \textbf{T.~Lindner:} conceptualization (Major (Equal)); data curation (Major (Equal)); funding acquisition (Minor (Supporting)); investigation (Minor (Supporting)); resources (Minor (Supporting)); software (Major (Equal)). \textbf{S.~Longo:} funding acquisition (Major (Equal)); investigation (Minor (Supporting)); project administration (Minor (Supporting)); supervision (Minor (Supporting)); writing – review \& editing (Minor (Supporting)). \textbf{K.W.~Madison:} funding acquisition (Major (Equal)); writing – review \& editing (Minor (Supporting)). \textbf{J.~Malcolm:} investigation (Minor (Supporting)); resources (Minor (Supporting)). \textbf{J.~Mammei:} funding acquisition (Major (Equal)). \textbf{R.~Mammei:} conceptualization (Minor (Supporting)); funding acquisition (Major (Equal)); investigation (Minor (Supporting)); project administration (Lead); resources (Lead); validation (Major (Equal)). \textbf{C.~Marshall:} conceptualization (Lead); resources (Lead). \textbf{J.~Martin:} conceptualization (Lead); funding acquisition (Major (Equal)); investigation (Minor (Supporting)); methodology (Lead); project administration (Lead); resources (Lead); supervision (Lead); writing – review \& editing (Major (Equal)). \textbf{R.~Matsumiya:} investigation (Major (Equal)); resources (Major (Equal)). \textbf{M.~McCrea:} investigation (Minor (Supporting)). \textbf{E.~Miller:} formal analysis (Major (Equal)); investigation (Major (Equal)); project administration (Minor (Supporting)); resources (Minor (Supporting)); writing – review \& editing (Major (Equal)). \textbf{M.~Miller:} investigation (Minor (Supporting)). \textbf{K.~Mishima:} conceptualization (Major (Equal)); data curation (Minor (Supporting)); funding acquisition (Major (Equal)); project administration (Lead); resources (Major (Equal)); software (Minor (Supporting)); supervision (Minor (Supporting)); writing – review \& editing (Major (Equal)). \textbf{T.~Mohammadi:} investigation (Minor (Supporting)). \textbf{T.~Momose:} funding acquisition (Major (Equal)). \textbf{M.~Nalbandian:} investigation (Minor (Supporting)); resources (Minor (Supporting)). \textbf{T.~Okamura:} funding acquisition (Major (Equal)); methodology (Major (Equal)); resources (Major (Equal)). \textbf{S.~Pankratz:} investigation (Minor (Supporting)). \textbf{R.~Patni:} investigation (Minor (Supporting)). \textbf{R.~Picker:} conceptualization (Lead); formal analysis (Major (Equal)); funding acquisition (Major (Equal)); investigation (Lead); methodology (Major (Equal)); project administration (Major (Equal)); resources (Lead); software (Minor (Supporting)); supervision (Lead); visualization (Lead); writing – original draft (Lead); writing – review \& editing (Lead). \textbf{V.~Purcell:} resources (Minor (Supporting)). \textbf{K.~Qiao:} investigation (Minor (Supporting)). \textbf{W.~Rathnakela:} methodology (Minor (Supporting)). \textbf{Y.-N.~Rao:} conceptualization (Minor (Supporting)); investigation (Major (Equal)); methodology (Major (Equal)); resources (Minor (Supporting)); validation (Minor (Supporting)). \textbf{T.~Reimer:} investigation (Minor (Supporting)). \textbf{D.~Salazar:} investigation (Minor (Supporting)). \textbf{W.~Schreyer:} conceptualization (Lead); data curation (Major (Equal)); investigation (Major (Equal)); methodology (Lead); resources (Major (Equal)); software (Major (Equal)); writing – review \& editing (Major (Equal)). \textbf{T.~Shima:} funding acquisition (Major (Equal)). \textbf{H.M.~Shimizu:} funding acquisition (Major (Equal)). \textbf{S.~Siddiqui:} resources (Minor (Supporting)). \textbf{S.~Sidhu:} investigation (Minor (Supporting)); methodology (Minor (Supporting)); resources (Minor (Supporting)); software (Minor (Supporting)). \textbf{S.~Stargardter:} investigation (Major (Equal)); resources (Major (Equal)). \textbf{R.~Stutters:} investigation (Minor (Supporting)); resources (Minor (Supporting)). \textbf{P.~Switzer:} resources (Minor (Supporting)). \textbf{Tushar:} investigation (Minor (Supporting)); resources (Minor (Supporting)). \textbf{M.~Uzair:} resources (Minor (Supporting)). \textbf{S.~Vanbergen:} conceptualization (Minor (Supporting)); data curation (Minor (Supporting)); formal analysis (Minor (Supporting)); investigation (Lead); methodology (Major (Equal)); project administration (Minor (Supporting)); resources (Lead); software (Major (Equal)); supervision (Minor (Supporting)); writing – review \& editing (Major (Equal)). \textbf{W.T.H.~van~Oers:} conceptualization (Minor (Supporting)); funding acquisition (Major (Equal)); supervision (Minor (Supporting)). \textbf{N.~Yazdandoost:} formal analysis (Minor (Supporting)); funding acquisition (Major (Equal)); investigation (Lead); project administration (Minor (Supporting)); resources (Minor (Supporting)); supervision (Minor (Supporting)); writing – original draft (Minor (Supporting)); writing – review \& editing (Major (Equal)). \textbf{Q.~Ye:} investigation (Minor (Supporting)); resources (Minor (Supporting)). \textbf{A.~Zahra:} investigation (Minor (Supporting)). \textbf{L.~Zhang:} data curation (Minor (Supporting)); methodology (Major (Equal)); resources (Minor (Supporting)); software (Major (Equal)).

%Conceptualization: A.B., C.D., and E.F. 
%Methodology \& Software: C.D. and G.H. 
%Formal Analysis: A.B., I.J., and K.L. 
%Investigation (Experiment/Simulation): G.H., I.J., M.N., and O.P. 
%Data Curation: K.L. and M.N. 
%Visualization: A.B. and I.J. 
%Supervision \& Project Administration: E.F. and Q.R. 
%Funding Acquisition: E.F. 
%Writing – Original Draft Preparation: A.B. and C.D. 
%Writing – Review \& Editing: All authors contributed to the discussion of the results and commented on the final manuscript.

% ==========================================
% 2. BIBLIOGRAPHY (Core word count stops before this)
% ==========================================
\bibliographystyle{apsrev4-2}
\bibliography{UCN_PRL_updated}
%\bibliography{UCNbib_PRL_cleaned}% Produces the bibliography via BibTeX.

% ==========================================
% 3. END MATTER START
% ==========================================
\appendix % This flag shifts everything below it into the End Matter zone

%\textcolor{purple}{End matter below can be up to two pages: \url{https://journals.aps.org/prl/authors#style-and-formatting}.} \\
%\noindent
{\it \textbf{Uncertainty determination}}---
%\section{Uncertainty determination}
\label{app:uncertainty}

{\it UCN counts and count rates}---
%\subsection{UCN counts}
\label{app:UCNcountuncertainty}
UCN counts in the lithium-glass detector are isolated from electronic noise and \(\gamma \)-ray backgrounds using pulse shape discrimination techniques detailed in Ref.~\cite{Jamieson17}.
For \emph{batch production and storage lifetime experiments}, residual backgrounds are quantified and subtracted using dedicated closed-valve cycles as baselines. Time-varying backgrounds (e.g. activation) are negligible, uncertainties are evaluated via Poisson statistics and Gaussian error propagation.
%Any residual background events for \emph{batch production and storage lifetime experiments} are quantified and subtracted by running dedicated experimental cycles with the gate valves closed, which serve as a baseline for the open-valve production cycles.
%Additional time-varying backgrounds such as activation were found to be negligible.
%The final uncertainties plotted for these measurements are calculated using Poisson statistics combined with Gaussian error propagation.
\emph{Continuous UCN production and detection experiments} exhibit fast and slow count rate decay during beam-off periods (inset, Fig.~\ref{fig:SSP}).
%\emph{Continuous UCN production} and detection experiments exhibit a fast and a slow decay of the count rate during the beam-off period as visible in the inset of Figure~\ref{fig:SSP}.
The fast component reflects UCN drainage through the guide system, while the slow component is attributed to UCN-induced activation of the detector:
\[
^{27}_{13}\text{Al} + {}^{1}_{0}\text{n} \rightarrow {}^{28}_{13}\text{Al} + \gamma.
\]
UCN can reach the aluminum housing of the detector via a roughly \qty{3}{\milli \meter} wide slit between the lithium glass and the detector top flange.
The subsequent $\beta^-$-decay 
\[
^{28}_{13}\text{Al} \rightarrow {}^{28}_{14}\text{Si} + e^- + \bar{\nu}_e
\] 
has a lifetime of \qty{194.3(2)}{\second} and a Q value of \qty{4.64226(12)}{\mega \electronvolt}. The decay electrons can reach the acrylic light guide inside the detector housing, inducing Cherenkov radiation.
The daughter nucleus $^{28}_{14}\text{Si}$ emits a \qty{1.778987(15)}{\mega \electronvolt} gamma ray~\cite{Al28IAEA}, which Compton scatters in the light guide;
the resulting free electrons also produce Cherenkov light.
These Cherenkov signals fall within the pulse-shape discrimination window used for UCN detection~\cite{Jamieson17}.
%An evaluation of the slow decay shows that the relative contribution to the count rate during irradition is smaller than \qty{2(1)e3}.
To account for this slow decay, we fit each UCN count rate histogram with a single exponential decay plus background $N_{\rm BG}(t) = N_0 e^{-t/\tau_{\rm BG}}+C_{\rm BG}$ in the interval [\qty{200}{\second}, \qty{800}{\second}] after the irradiation.
The extracted decay constants align reasonably well with $^{28}_{13}\text{Al}$ decay.
Extrapolating \(N_{\rm BG}(t)\) to the end of the irradiation period yields an estimate for the saturated background which we subtract from the data.
This background constitutes \(< 1\%\) of the saturated count rate for all points in Fig.~\ref{fig:SSP}, with uncertainties propagated from the fit parameters.

{\it Buffer overrun correction}---
\begin{figure}[htb]
%trim=left bottom right top
\centering \includegraphics[width=\columnwidth, page=1,trim=0cm 0cm 0cm 0cm,clip]{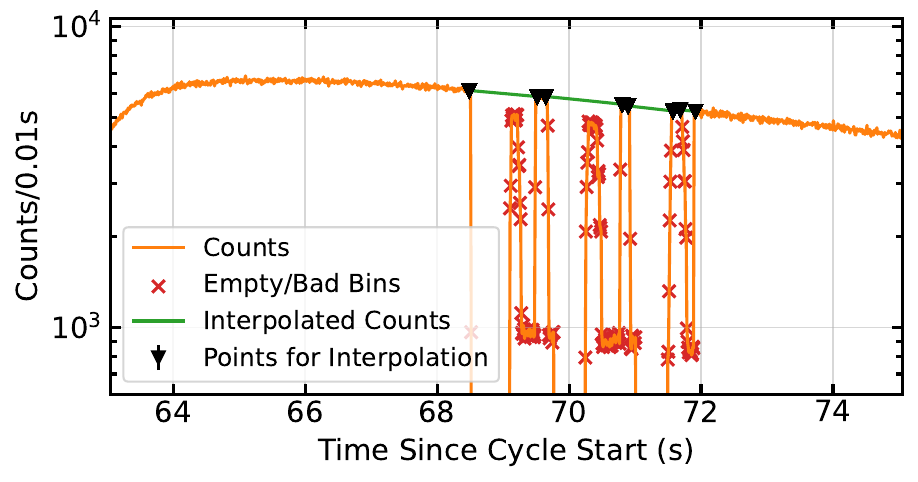}
\caption{Time histogram of UCN events in the DAQ system showing dropped counts due to buffer overrun as red crosses. The green lines are linearly interpolated between the bins denoted by black triangles. This run corresponds to a period with gate valves open to the detector of the batch production run without foil at a proton current of \qty{26.6(3)}{\micro \ampere}. The number of UCN detected increased from \qty{1.34(1)e7} to $1.47(2)\times 10^7$ due to the correction.
\label{fig:buffercorrection} }
\end{figure}
%\subsection{Beamline 1U current}
\label{app:buffer}
For the data points at highest currents and without foil in Figs.~\ref{fig:batch} and \ref{fig:SSP}, the rates in the $^6$Li detector exceeded the transmission capability of our data acquisition system:
above around \qty{6e5} events per second, the buffer in the CAEN V1725 digitizer runs full and events are dropped, as shown in Fig.~\ref{fig:buffercorrection} as an example for batch production.
We performed a simple linear interpolation using the five unaffected bins adjacent to the intervals with dropped events.
As a validation, the same regions were reconstructed using exponential fits to the surrounding emptying curve, yielding integrated counts that differed from the linear-interpolation result by at most 0.44\%.
Additional tests were performed on unaffected cycles by introducing artificial gaps with the same timing and duration and reconstructing them with the same procedure.
These tests reproduced the known integrated counts to within 0.17–0.35\%, with the reconstructed values lying slightly below the original counts.
To account for this, we quadratically added a fractional uncertainty of 0.44\% to the integrated counts of the affected experimental cycles.

For the steady state production data point without foil at \qty{26.6(3)}{\micro \ampere} in Fig.~\ref{fig:SSP} dropped events affected a significant number of bins.
These bins were excluded for the exponential saturation curve fit.

{\it Pileup}---
\label{app:pileup}
Our lithium-glass detector is segmented into nine pixels to handle large event rates.
However, rates approaching \qty{1e6}{\per \second} that we saw for high proton currents could lead to neutron-neutron pileup becoming significant.
It shows up at the top right in the pulse shape discrimination (PSD) plot, as described in Ref.~\cite{Jamieson17}.
By cutting these regions, we determined that pileup constitutes less than 0.5\% of the UCN counts for the highest current (\qty{33.3(3)}{\micro \ampere}) in batch production.
Given source fluctuations easily exceed several percent between cycles, we neglect pileup for this study.

{\it Beamline 1U current}---
%\subsection{Beamline 1U current}
\label{app:BL1Uuncertainty}
The proton current injected into beamline 1 is derived from the electron current measured at the cyclotron’s extraction foil and foil holder, where a fraction of the two electrons stripped from the circulating H$^-$ ions are captured. This extraction foil system is calibrated annually against a toroidal, non-intercepting current monitor (TNIM) located in beamline 1~\cite{rawnsley1995beam, rawnsley2013beam}.
The TNIM is itself calibrated annually using an integrated, in-situ calibration wire to inject a reference current.
Through this calibration chain, the overall relative uncertainty of the primary current measurement is estimated to be approximately 1\%.
To feed beamline 1U (BL1U), a fast kicker diverts exactly one out of an adjustable number of proton pulses from the main cyclotron beam~\cite{bib:kicker} away from beamline 1.
The resulting BL1U current is calculated by multiplying the primary beamline 1 current by this known kick fraction.
These values are recorded at a 1~Hz sampling rate;
averaging over the irradiation period of the neutron spallation target yields the final BL1U current, with the standard deviation of the dataset defining the experimental error bars.

{\it Fill volume of deuterium moderator vessel}---
%\subsection{Fill volume of deuterium moderator vessel}
\label{app:D2volume}
Severe radiation and spatial constraints preclude direct level measurements within the liquid deuterium moderator vessel. 
We use a multi-step process to determine the LD$_2$ fill volumes and levels in Fig.~\ref{fig:batchvsD2level} from pressures and temperatures.
(1) The mass of deuterium recovered into the storage tank, \(\Delta m\), is determined from thermo-physical properties using CoolProp~\cite{bell2014coolprop} as \(\Delta m = m(\Delta P_{\rm tank}, T_{\rm tank}, V_{\rm tank})\), where \(\Delta P_{\rm tank}\) is the measured pressure change during recovery, and \(T_{\mathrm{t}ank}\) and \(V_{\mathrm{t}ank}\) are the tank temperature and volume, respectively. 
(2) The remaining liquid volume is then computed as \(V_{\rm LD2} = V_{\rm vessel} - \Delta V(P_{\rm sat},\Delta m)\), where \(V_{\rm vessel}\) is the calibrated moderator vessel volume, and \(P_{\mathrm{s}at}\), the saturated vapor pressure measured above the cryostat, determines the liquid density. 
(3) Since \(T_{\mathrm{t}ank}\) is inferred from local meteorological data, it introduces significant uncertainty. 
To quantify this, we correlate tank pressure and temperature fluctuations during a benchmark period of constant deuterium content post-UCN data acquisition.
The uncertainty on \(T_{\mathrm{t}ank}\) is bounded by the maximum observed pressure spread during ambient temperature shifts equivalent to those experienced during the recovery phase used for Fig.~\ref{fig:batchvsD2level}.

\noindent
{\it \textbf{Storage-lifetime fit parameters}}---
\label{app:fitparams}
%The fit parameters from Fig.~\ref{fig:SSL} can be found in 
Table~\ref{tab:table1} shows the parameters of the bi-exponential fit for the storage lifetime measurement of UCN inside the TUCAN source (Fig.~\ref{fig:SSL}).
The reduced $\chi^2$ in the last column is quite large indicating possible unaccounted systematic effects or source fluctuations.
%The spread of fitted parameters between beam-current settings exceeds the formal statistical fit uncertainties.
\begin{table}[h!]
    \caption{Fit parameters for Fig.~\ref{fig:SSL}}
    \label{tab:table1}
    \begin{tabular}{c|c|c|c|c|c}
      \shortstack{Current\\($\mu$A)}& $N_0\,(\times10^5)$ & $\tau_1$ (s) & $\tau_2$ (s) & $f$ & \raisebox{0.5ex}{$\frac{\chi^2}{DOF}$}\\
      \hline
      %1.00(1)&$1.921(8)$&$19.09(15)$& $41.65(33)$& $0.902(4)$&$2.847$\\
      1.00(1)&$1.921(8)$&$19.1(2)$& $41.7(3)$& $0.902(4)$&$2.8$\\
      %5.00(5)&$9.764(37)$&$19.71(10)$& $41.55(15)$& $0.920(2)$& $5.581$\\
      5.00(5)&$9.76(4)$&$19.7(1)$& $41.6(2)$& $0.920(2)$& $5.6$\\      
      %10.0(1)&$19.709(73)$&$19.98(9)$& $41.85(18)$&$0.927(1)$&$5.001$\\
      10.0(1)&$19.71(7)$&$20.0(1)$& $41.9(2)$&$0.927(1)$&$5.0$\\
      %20.0(2)&$38.97(14)$&$20.40(8)$& $42.67(26)$&$0.933(1)$&$5.067$\\
      20.0(2)&$39.0(1)$&$20.4(1)$& $42.7(3)$&$0.933(1)$&$5.1$\\
      %33.0(3)&$64.38(22)$&$20.60(8)$& $42.84(16)$&$0.929(1)$&$8.522$\\
      33.0(3)&$64.4(2)$&$20.6(1)$& $42.8(2)$&$0.929(1)$&$8.5$\\
    \end{tabular}
\end{table}

\end{document}